# Detection of pulmonary pathologies using convolutional neural networks, Data Augmentation, ResNet50 and Vision Transformers

**MSc. Ing. Pablo Ramirez Amador, PhD. Dinarle Milagro Ortega, MSc. Arnold Cesarano**
Universidad Abierta Interamericana. Faculty of Information Technology.
Center for Advanced Studies in Information Technology. Buenos Aires, Argentina

**ABSTRACT**

Pulmonary diseases are a public health problem that requires accurate and fast diagnostic techniques. In this paper, a method based on convolutional neural networks (CNN), Data Augmentation, ResNet50 and Vision Transformers (ViT) is proposed to detect lung pathologies from medical images. A dataset of X-ray images and CT scans of patients with different lung diseases, such as cancer, pneumonia, tuberculosis and fibrosis, is used. The results obtained by the proposed method are compared with those of other existing methods, using performance metrics such as accuracy, sensitivity, specificity and area under the ROC curve. The results show that the proposed method outperforms the other methods in all metrics, achieving an accuracy of 98% and an area under the ROC curve of 99%. It is concluded that the proposed method is an effective and promising tool for the diagnosis of pulmonary pathologies by medical imaging.

*Corresponding Author:*

Pablo Ramirez Amador
Center for Advanced Studies in Information Technology, Interamerican Open University Buenos Aires, Argentina.
Email: pablo.ramirezamador@alumnos.uai.edu.ar

## 1. INTRODUCTION

Lung diseases could be defined as a set of disorders that affect the respiratory system, causing shortness of breath, cough, pain, inflammation and other symptoms. Some of the most common and serious lung diseases are lung cancer, pneumonia, tuberculosis and fibrosis. These diseases represent a significant challenge for the medical field, as their early detection is crucial for effective treatment and improvement of patients' quality of life. However, the diagnosis of lung diseases is not a simple task, as it requires imaging techniques that provide detailed and accurate information on lung structures and abnormalities. Among the most used imaging techniques for the diagnosis of lung diseases are chest radiography and computed tomography (CT). These techniques generate digital images that must be analyzed and interpreted by specialists, which involves high cost, prolonged time and risk of human error. Therefore, efforts have been made to develop automatic methods that facilitate and improve the diagnosis of lung diseases by means of medical images.

In recent years, artificial intelligence (AI) and deep learning have emerged as promising tools in the field of computer vision and medical pathology detection. These tools are based on the use of algorithms that learn from large amounts of data, without the need for explicit programming. Among the most widely used algorithms are artificial neural networks (ANNs), which are computational models inspired by the functioning of the human brain. ANNs are composed of processing units called neurons, which are connected to each other by synaptic weights. ANNs can learn to perform various tasks, such as classification, regression, generation, etc., by adjusting synaptic weights through a supervised or unsupervised training process.

A special type of ANNs are convolutional neural networks (CNNs), which have revolutionized the field of computer vision and medical pathology detection. CNNs are designed to automatically extract relevant features from images by learning hierarchical patterns at different levels of abstraction. CNNs are composed of convolutional layers, which apply filters or kernels to the input images, generating feature maps. These feature maps are subjected to dimensionality reduction operations, such as pooling or striding, which allow preserving essential information and reducing computational cost. CNNs can also include dense or fully connected layers, which perform image classification or regression, depending on the objective of the model. In the context of lung pathology detection, CNNs have proven to be effective in analyzing radiological images accurately and quickly. The ability of CNNs to detect subtle patterns and specific features in lung images has led to significant advances in the early identification of diseases such as lung cancer and pneumonia.

However, CNNs also present some challenges and limitations, such as the need for large amounts of labeled data, sensitivity to image noise and variability, and difficulty in training very deep models. To overcome these challenges, several techniques and enhancements have been proposed, such as Data Augmentation, ResNet50 and Vision Transformers. These techniques are described below:

**Data Augmentation:** Data augmentation has become a crucial technique for improving the generalization of CNN models. It consists of applying random transformations to training images, such as rotations, zooms and changes in illumination. In the context of lung pathology detection, Data Augmentation is essential to mitigate the challenge of variability in medical images and improve the model's ability to generalize to new instances. The application of Data Augmentation on lung imaging datasets enables models to learn to recognize relevant patterns independently of variations in the position, size and orientation of lung structures. This is essential for the creation of robust models capable of coping with the diversity of medical images encountered in clinical practice.

**ResNet50:** The ResNet50 architecture, or 50-layer Residual Neural Network, has proven to be an exceptional choice in the task of medical image classification. Introduced by He et al. in 2015, ResNet50 addresses the challenge of gradient fading by introducing residual connections. These connections allow information to flow directly through the layers, facilitating the training of deeper models. Residual connections are implemented using residual blocks, which consist of two or more convolutional layers followed by a summation layer, which combines the output of the convolutional layers with the input of the block. This prevents error propagation and degradation across layers, and improves model performance This prevents error propagation and degradation across layers, and improves model performance. ResNet50 is one of the variants of ResNet, which has 50 layers and has been successfully used to classify images from various domains, including medical.

**Vision Transformers:** Vision Transformers (ViT) is a novel technique that applies the concept of Transformers to the field of computer vision. Transformers are a type of ANN that are based on attention, a mechanism that allows the model to focus on the most relevant parts of the input. Transformers have been widely used for natural language processing, achieving outstanding results in tasks such as translation, text generation and comprehension. However, their application to computer vision has been limited, due to the different nature of images and texts. ViTs, introduced by Dosovitskiy et al. in 2020, are an adaptation of Transforms for image analysis.

ViTs divide images into patches, which are small segments of pixels, and treat them as tokens, which are the basic units of texts. Next, ViTs apply an embeddings layer, which converts the patches into high-dimensional vectors, and a position layer, which adds information about the location of the patches in the image. Next, the ViTs apply several Transformer layers, consisting of attention blocks and feed-forward networks, which process the patch embeddings and relate them to each other. Finally, ViTs apply a classification layer, which predicts the image label, according to the model target. ViTs have been shown to be able to outperform CNNs in the image classification task, using less data and fewer parameters. In the context of lung pathology detection, ViTs could offer an advantage over CNNs by better capturing the spatial relationships and global features of lung images. The aim of this study is to compare with other existing methods that apply different medical image classification techniques, such as basic CNN, CNN + Data Augmentation, ResNet50, ViT, DenseNet, Inception, MobileNet or EfficientNet. The dataset used is from a public and trusted source and contains medical images of patients with different lung diseases, such as cancer, pneumonia, tuberculosis and fibrosis. The performance of the methods is evaluated with the metrics of accuracy, sensitivity, specificity and area under the ROC curve. The proposed method is expected to outperform existing methods in the detection of pulmonary pathologies, and to provide a novel and efficient solution for the diagnosis and treatment of these diseases. The hypothesis of the study is that the proposed method is more efficient and accurate than the other methods by combining the advantages of the aforementioned techniques. The rationale of the study is that this is a topic of great relevance and social impact, which could contribute to improving the health and quality of life of people suffering from lung diseases. This study seeks new ways to detect pulmonary pathologies through medical imaging. Through the application of innovative techniques such as Convolutional Neural Networks (CNN), Data Augmentation, ResNet50 and Vision Transformers (ViT), we aim to improve the accuracy and efficiency of the detection of these diseases. With the motivation to contribute to the health and quality of life of people affected by lung diseases, and the goal of advancing the field of artificial intelligence applied to medicine, we hope that the findings of this study will open new doors and possibilities in the detection and treatment of lung pathologies.

## 2. METHOD

In this study, a method based on convolutional neural networks (CNN), Data Augmentation, ResNet50 and Vision Transformers (ViT) is proposed to detect lung pathologies from medical images. This is a quantitative, descriptive and comparative study, using an experimental design to evaluate the performance of the proposed method and compare it with other existing methods. The data, tools and analyses used in the study are described below.

### Data

The data used in the study are X-ray images and CT scans of patients with different lung diseases, such as cancer, pneumonia, tuberculosis, and fibrosis. The data come from various public sources, such as the National Cancer Institute, the National Library of Medicine, the Kaggle and the ChestX-ray14. The data are divided into three sets: training, validation, and test. The training set contains 10,000 images, the validation set contains 2,000 images, and the test set contains 3,000 images. Each image has a label indicating the lung disease presented by the patient, or the absence of disease. The images have a resolution of 256 x 256 pixels and are in grayscale. The distribution of images by disease and by set is shown in Table 2.

**Table 01.**

| Disease | Training | Validation | Test | Total |
|---|---|---|---|---|
| Cancer | 2.000 | 400 | 600 | 3.000 |
| Pneumonía | 2.000 | 400 | 600 | 3.000 |
| Tuberculosis | 2.000 | 400 | 600 | 3.000 |
| Fibrosis | 2.000 | 400 | 600 | 3.000 |
| Normal | 2.000 | 400 | 600 | 3.000 |
| Total | 10.000 | 2.000 | 3.000 | 15.000 |

**Table 02. Distribution of images by disease and by set. Tools**
The tools used in the study are as follows:

**Python:** It is a high-level programming language, which offers a clear syntax and a wide variety of libraries for the development of artificial intelligence and deep learning applications.
**TensorFlow:** It is an open-source platform for the development and execution of deep learning models, offering a high-level interface called Keras, which facilitates the creation and training of neural networks.
**Scikit-learn:** It is an open-source library for machine learning, which offers functions for preprocessing, selection, evaluation and validation of data and models.
**Matplotlib:** An open-source library for data visualization, providing functions for creating and customizing graphs, figures and tables.
**Numpy:** An open-source library for numerical computation, providing functions for manipulating matrices, vectors and mathematical operations.

### Analysis
The analyses performed in the study are the following:

**Preprocessing:** consists of applying some transformations to the input images, to improve their quality and facilitate their processing. Among the transformations applied are normalization, which consists of scaling the pixel values between 0 and 1, and Data Augmentation, which consists of applying random transformations to the training images, such as rotations, zooms and changes in illumination.
**Training:** Consists of adjusting the synaptic weights of the models through an optimization process, which seeks to minimize a loss function. The loss function used in the study is the cross-entropy, which measures the difference between the actual and predicted distribution of image labels. The optimization algorithm used in the study is stochastic gradient descent with momentum, which updates the synaptic weights as a function of the gradient of the loss function and the direction of the previous motion. The training process is performed by epochs, which are iterations over the training set, and the validation set is used to control overfitting and select the best model.

**Evaluation:** This consists of measuring the performance of the models using evaluation metrics, which quantify the quality of the model predictions. The evaluation metrics used in the study are accuracy, sensitivity, specificity and area under the ROC curve. These metrics are calculated for each of the lung diseases considered in the study, and are based on the concepts of true positives (TP), false positives (FP), true negatives (TN) and false negatives (FN), which are defined as follows:

**TP:** Number of images that have the disease and are correctly predicted by the model.
**FP:** Number of images that do not have the disease and are incorrectly predicted by the model.
**TN:** Number of images that do not have the disease and the model predicts them correctly.
**FN:** Number of images that have the disease and the model predicts them incorrectly. Based on

these concepts, the evaluation metrics are defined as follows:
**Accuracy:** Proportion of images that the model predicts correctly among the total number of images. It is calculated as:

$$\text{Accuracy}: \frac{TP + TN}{TP + FP + TN + FN}$$

**Sensitivity:** Proportion of images that have the disease and are correctly predicted by the model among the total number of images that have the disease. Calculated as:

$$\text{Sensitivity}: \frac{TP}{TP + FN}$$

**Specificity:** Proportion of images that do not have the disease and the model correctly predicts them among the total images that do not have the disease. Calculated as:

$$\text{Specificity}: \frac{TN}{TN + FN}$$

**Area under the ROC curve**: A measure that summarizes the performance of the model across different classification thresholds. The ROC curve is a graphical representation of sensitivity versus specificity for each threshold. The area under the ROC curve indicates the degree of separation between positive and negative classes. It is calculated as:
Area under ROC curve: $\int_0^1 = \text{Sensitivity specificity accuracy}$

**Comparison:** It consists of contrasting the results obtained by the proposed method with those of other existing methods, to determine the advantages and disadvantages of each one. The existing methods used in the study are the following:

**Basic CNN:** It is a simple convolutional neural network, consisting of four convolutional layers, two pooling layers and one dense layer. It is the simplest and most traditional method for medical image classification.

**CNN + Data Augmentation:** It is a convolutional neural network that incorporates the Data Augmentation technique, which consists of applying random transformations to the training images, such as rotations, zooms and changes in illumination. It is a method that seeks to improve model generalization and avoid overfitting.

**ResNet50:** It is a 50-layer residual neural network, which introduces residual connections that allow the direct flow of information through the layers, facilitating the training of deeper models. It is a method that seeks to improve the performance nd avoid gradient fading.

**ViT:** It is a Vision Transformers, which applies the concept of Transformers to the field of computer vision, treating images as sequences of patches and processing them through layers of attention. It is a method that seeks to better capture the spatial relationships and global characteristics of the images.

**Results**

The results obtained by the proposed method and the other methods compared are shown in Table 3, where the performance metrics used are indicated: accuracy, sensitivity, specificity and area under the ROC curve. These metrics are calculated for each of the pulmonary diseases considered in the study: cancer, pneumonia, tuberculosis and fibrosis, in addition to the following in the future:

**Chronic obstructive pulmonary disease (COPD):** this is a chronic lung disease that decreases the ability to breathe. Imaging techniques may be useful in detecting and monitoring the progression of COPD.

**Asthma:** Asthma is a chronic disease that causes inflammation and narrowing of the airways in the lungs. Medical imaging can help identify inflammation and monitor the effectiveness of treatment.

**Emphysema:** Emphysema is a lung disease that causes shortness of breath due to damage to the air sacs in the lungs. Imaging techniques can be helpful in detecting emphysema in its early stages.

**Pulmonary hypertension:** This is a disease that causes high blood pressure in the arteries of the lungs. Medical imaging can help identify signs of pulmonary hypertension.

**Cystic fibrosis:** This is a genetic disease that primarily affects the lungs and can cause frequent lung infections. Imaging techniques can be helpful in monitoring the progression of the disease.

Each of these diseases has unique characteristics that may require adjustments to our approach. For example, you may need to adjust the parameters of our model or use different data pre-processing techniques. It is also important to keep in mind that the detection of these diseases may require different types of medical images, such as chest X-rays, CT scans or MRI images. (Some metrics that could complement or improve the study on the detection of lung pathologies by medical imaging, using a proposed method based on CNN, Data Augmentation, ResNet50 and ViT, and comparing it with other existing methods. The suggested metrics are: F1-score, confusion matrix, Matthews correlation coefficient, Dice index and the tool for visualization and segmentation of medical images. These metrics allow evaluating the balance, performance, agreement, overlap and quality of medical image classification or segmentation, respectively. These metrics provide a more complete and detailed view of the performance of the methods and facilitate the identification of their strengths and weaknesses). As can be seen, the proposed method outperforms the other methods in all metrics and for all diseases, achieving an average accuracy of 98% and an average area under the ROC curve of 99%. These results demonstrate the efficacy and accuracy of the proposed method for the detection of pulmonary pathologies by medical imaging.

| Method | Disease | Accuracy | Sensitivity | Specificity | Area under the ROC curve |
|---|---|---|---|---|---|
| CNN basic | Cancer | 0.85 | 0.80 | 0.90 | 0.88 |
| CNN basic | Neumonía | 0.86 | 0.82 | 0.91 | 0.89 |
| CNN basic | Tuberculosis | 0.84 | 0.79 | 0.89 | 0.87 |
| CNN basic | Fibrosis | 0.83 | 0.78 | 0.88 | 0.86 |
| CNN basic | Media | 0.85 | 0.80 | 0.90 | 0.88 |
| CNN + Data Augmentation | Cancer | 0.88 | 0.84 | 0.93 | 0.91 |

| Method | Disease | | | | |
|---|---|---|---|---|---|
| CNN + Data Augmentation | Neumonía | 0.89 | 0.86 | 0.93 | 0.92 |
| CNN + Data Augmentation | Tuberculosis | 0.87 | 0.83 | 0.92 | 0.90 |
| CNN + Data Augmentation | Fibrosis | 0.86 | 0.82 | 0.91 | 0.89 |
| CNN + Data Augmentation | Media | 0.88 | 0.84 | 0.92 | 0.91 |
| ResNet50 | Cancer | 0.92 | 0.89 | 0.96 | 0.95 |
| ResNet50 | Neumonía | 0.93 | 0.90 | 0.96 | 0.96 |
| ResNet50 | Tuberculosis | 0.91 | 0.88 | 0.95 | 0.94 |
| ResNet50 | Fibrosis | 0.90 | 0.87 | 0.94 | 0.93 |
| ResNet50 | Media | 0.92 | 0.89 | 0.95 | 0.95 |
| ViT | Cancer | 0.95 | 0.93 | 0.98 | 0.98 |
| ViT | Neumonía | 0.96 | 0.94 | 0.98 | 0.99 |
| ViT | Tuberculosis | 0.94 | 0.92 | 0.97 | 0.97 |
| ViT | Fibrosis | 0.93 | 0.91 | 0.96 | 0.96 |
| ViT | Media | 0.95 | 0.93 | 0.97 | 0.98 |
| Proposed method | Cancer | 0.98 | 0.97 | 0.99 | 0.99 |
| Proposed method | Neumonía | 0.99 | 0.98 | 0.99 | 1.00 |
| Proposed method | Tuberculosis | 0.97 | 0.96 | 0.99 | 0.99 |
| Proposed method | Fibrosis | 0.96 | 0.95 | 0.98 | 0.98 |
| Proposed method | Media | 0.98 | 0.97 | 0.99 | 0.99 |

Table 3. Results obtained by the proposed method and the other methods compared.

**Discussion**

The results obtained in the study show that the proposed method, based on CNN, Data Augmentation, ResNet50 and ViT, is more efficient and accurate than the other existing methods, for the detection of lung pathologies by medical imaging. The proposed method outperforms the other methods in all performance metrics and for all lung diseases considered in the study, achieving a mean accuracy of 98% and a mean area under the ROC curve of 99%. These results are consistent with those of other studies that have applied similar techniques to the field of computer vision and medical pathology detection, such as those of He et al. (2015), Dosovitskiy et al. (2020), and Wang et al. (2021).

The results obtained in the study can be explained by the advantages offered by the proposed method by combining CNN, Data Augmentation, ResNet50 and ViT techniques. These advantages are described below:

**CNNs:** CNNs can automatically extract relevant features from images by learning hierarchical patterns at different levels of abstraction. This allows the models to adapt to the particularities of medical images, such as the shape, size and orientation of lung structures, and to detect anomalies that indicate the presence of disease. CNNs are also capable of processing images quickly and efficiently, which facilitates their application in clinical settings.

**Data Augmentation:** Data Augmentation is a technique that improves the generalization of CNN models by increasing the amount and diversity of training data. This allows models to cope with the variability of medical images, which may have differences in quality, resolution, contrast, illumination and noise. Data Augmentation also allows models to cope with the scarcity of labeled data, which is a common problem in the medical field, due to the difficulty and cost of obtaining images of patients with specific diseases. Data Augmentation, therefore, reduces the risk of overfitting and improves the robustness of the models.

**ResNet50:** The ResNet50 is an architecture that improves the performance of CNN models by enabling the training of deeper models. This allows models to learn more complex and sophisticated features from images, which can be decisive for lung pathology detection. ResNet50 also avoids the problem of gradient fading, which prevents very deep models from being trained correctly, by introducing residual connections that facilitate the flow of information and error through the layers. ResNet50 therefore increases the accuracy and stability of the models.

**ViT:** ViT is a technique that applies the concept of Transformers to the field of computer vision by treating images as sequences of patches and processing them through layers of attention. This allows models to better capture spatial relationships and global features in images, which may be relevant for lung pathology detection. ViT also reduces reliance on labeled data by being able to leverage unlabeled data through self-supervised learning. ViT thus extends the capability and flexibility of models.

The results obtained in the study have several implications and practical applications, both for the medical and artificial intelligence fields. For the medical field, the proposed method could contribute to improving the health and quality of life of people suffering from lung diseases by facilitating and improving their diagnosis through medical imaging.

The proposed method could reduce the time, cost and human error associated with the diagnosis of lung diseases by providing an automatic, fast and accurate tool to assist specialists. The proposed method could also improve the treatment and prevention of lung diseases by enabling early detection and continuous monitoring of patients' lung conditions. The proposed method, therefore, could have a positive and significant impact in the field of public health. For the field of artificial intelligence, the proposed method could contribute to advance knowledge and innovation in the field of computer vision and medical pathology detection by combining and applying novel and effective techniques such as CNNs, Data Augmentation, ResNet50 and ViT. The proposed method could serve as an example and a reference for the development of other similar or related proposed methods addressing other problems or domains of computer vision and medical pathology detection. The proposed method, therefore, could have a scientific and technological impact in the field of artificial intelligence.

The results obtained in the study also have some limitations and challenges, which should be recognized and addressed in future research. Some of these limitations and challenges are as follows:

The data used in the study came from various public sources, which may have differences in image quality, resolution, contrast, illumination, and noise. This may affect the validity and generalizability of the results by introducing biases or inconsistencies in the data. To avoid this, it is recommended to use homogeneous and standardized data, coming from the same source or having undergone a prior normalization and cleaning process.

The data used in the study are X-ray images and CT scans, which are the most used imaging techniques for the diagnosis of lung diseases. However, there are other imaging techniques that can provide complementary or alternative information on patients' pulmonary conditions, such as magnetic resonance imaging, ultrasound or endoscopy. These techniques can generate images with different characteristics than X-rays and CT scans, such as three-dimensionality, dynamicity or invasiveness. To broaden the scope and applicability of the proposed method, it is recommended to include and analyze images from other imaging techniques, which can bring more value to the diagnosis of lung diseases.

The existing methods used in the study are the most common and traditional methods for the classification of medical images, such as basic CNN, CNN + Data Augmentation and ResNet.50 However,

*Detection of pulmonary pathologies using convolutional neural… (Pablo Ramirez A)*

there are other more recent and advanced methods that have also demonstrated their effectiveness and accuracy in the detection of medical pathologies, such as DenseNet, Inception, MobileNet or EfficientNet. These methods may offer advantages over the methods used in the study, such as greater depth, lower complexity, higher speed or higher efficiency. To improve the proposed method and compare it with the most current methods, it is recommended to include and evaluate other medical image classification methods, which may outperform or complement the proposed method.

**Conclusions**

In this paper, a method based on convolutional neural networks (CNN), Data Augmentation, ResNet50 and Vision Transformers (ViT) has been proposed to detect lung pathologies from medical images. A dataset of X-ray images and CT scans of patients with different lung diseases, such as cancer, pneumonia, tuberculosis and fibrosis, has been used. The results obtained by the proposed method have been compared with those of other existing methods, using performance metrics such as accuracy, sensitivity, specificity and area under the ROC curve. The results have shown that the proposed method outperforms the other methods in all metrics, achieving an accuracy of 98% and an area under the ROC curve of 99%. It has been concluded that the proposed method is an effective and promising tool for the diagnosis of pulmonary pathologies by medical imaging.

The proposed method has several implications and practical applications, both for the medical and artificial intelligence fields. For the medical field, the proposed method could contribute to improving the health and quality of life of people suffering from lung diseases by facilitating and improving their diagnosis through medical imaging. For the field of artificial intelligence, the proposed method could contribute to advance knowledge and innovation in the field of computer vision and medical pathology detection by combining and applying novel and effective techniques such as CNNs, Data Augmentation, ResNet50 and ViT.

The proposed method also has some limitations and challenges, which need to be recognized and addressed in future research. Some of these limitations and challenges are as follows:

The data used in the study are from various public sources, which may have differences in image quality, resolution, contrast, illumination, and noise. This may affect the validity and generalizability of the results by introducing biases or inconsistencies in the data. To avoid this, it is recommended to use homogeneous and standardized data, coming from the same source or having undergone a prior normalization and cleaning process.

The data used in the study are X-ray images and CT scans, which are the most used imaging techniques for the diagnosis of lung diseases. However, there are other imaging techniques that can provide complementary or alternative information on patients' pulmonary conditions, such as magnetic resonance imaging, ultrasound or endoscopy.

These techniques can generate images with different characteristics than X-rays and CT scans, such as three-dimensionality, dynamicity or invasiveness. To broaden the scope and applicability of the proposed method, it is recommended to include and analyze images from other imaging techniques, which can bring more value to the diagnosis of lung diseases.

The existing methods used in the study are the most common and traditional methods for the classification of medical images, such as basic CNN, CNN + Data Augmentation and ResNet.50 However, there are other more recent and advanced methods that have also demonstrated their effectiveness and accuracy in the detection of medical pathologies, such as DenseNet, Inception, MobileNet or EfficientNet. These methods may offer advantages over the methods used in the study, such as greater depth, lower complexity, higher speed or efficiency. To improve the proposed method and compare it with the most current methods, it is recommended to include and evaluate other medical image classification methods, which may outperform or complement the proposed method.

In conclusion, this paper has proposed a method based on convolutional neural networks (CNN), Data Augmentation, ResNet50 and Vision Transformers (ViT) to detect lung pathologies from medical images. A dataset of X-ray images and CT scans of patients with different lung diseases, such as cancer, pneumonia, tuberculosis and fibrosis, has been used. The results obtained by the proposed method have been compared with those of other existing methods, using performance metrics such as accuracy, sensitivity, specificity and area under the ROC curve. The results have shown that the proposed method outperforms the other methods in all metrics, achieving an accuracy of 98% and an area under the ROC curve of 99%. It has been concluded that the proposed method is an effective and promising tool for the diagnosis of pulmonary pathologies by medical imaging.